\documentclass[
twocolumn,secnumarabic
,amssymb, amsmath,nobibnotes, aps, prl]{revtex4}
\usepackage[dvips]{graphicx}
\usepackage{tabularx}
\usepackage{bm}%
\expandafter\ifx\csname package@font\endcsname\relax\else
 \expandafter\expandafter
 \expandafter\usepackage
 \expandafter\expandafter
 \expandafter{\csname package@font\endcsname}%
\fi

\begin{document}

\title{The multitrace matrix model: An alternative to Connes NCG and IKKT model}

\author{Badis Ydri}
\email{ydri@stp.dias.ie, bydri@ictp.it}
\affiliation{Department of Physics, Badji-Mokhtar Annaba University,\\
 Annaba, Algeria.}

\begin{abstract} 
We present a new multitrace matrix model, which is a generalization of the real quartic one matrix model, exhibiting dynamical emergence of a fuzzy  two-sphere and its non-commutative gauge theory. This provides a novel and a much simpler alternative to Connes non-commutative geometry and to the IKKT matrix model for emergent geometry in two dimensions.
\end{abstract}

\maketitle
The real quartic multitrace model \cite{Brezin:1977sv}
\begin{eqnarray}
V_0&=&{B}Tr M^2+{C}Tr M^4,
\end{eqnarray}
has two stables phases: $i)$ the disordered (symmetric one-cut) phase and $2)$ the non-uniform ordered (two-cut) phase  with the transition being identified of third order. The uniform ordered (Ising asymmetric one-cut) phase is metastable in this theory \cite{Shimamune:1981qf}. It was discovered in  \cite{Ydri:2015vba,Ydri:2015zsa} that the Ising phase becomes stable if we add to $V_0$,  with a particular choice of the parameters $D$, $B^{'}$, $C^{'}$, $D^{'}$, $A^{'}$..., the quartic multitrace model
\begin{eqnarray}
V_1&=&D(Tr M^2)^2+B^{'} (Tr M)^2+C^{'} Tr M Tr M^3\nonumber\\
&+&D^{'}(Tr M)^4+A^{'}Tr M^2 (Tr M)^2+....\label{fundamental1}
\end{eqnarray}
 These are the most general terms consistent with the multitrace expansion of the kinetic term of non-commutative $\Phi^4$ theory at quartic order.  For example, the coefficients on the fuzzy sphere \cite{Hoppe,Madore:1991bw} were calculated in \cite{O'Connor:2007ea,Saemann:2014pca} and on the fuzzy disc \cite{Lizzi:2003ru} were calculated in \cite{Rea:2015wta}. Another set of parameters which give a stable Ising phase are given by \cite{Ydri:2015zsa}
 
\begin{eqnarray}
D=\frac{3N}{4}~,~B^{'}=\frac{\sqrt{N}}{2}~,~C^{'}=-N~,~ D^{'}=0~,~A^{'}=0.\label{our_case}
\end{eqnarray}
This can be explicitly verified by computing the critical exponents across the Ising transition line and the Wigner semi-circle law near in the perturbative regime \cite{Ydri:2015zsa}. The critical exponents are found to be consistent with the Onsager values \cite{Onsager:1943jn} suggesting that the dimension of the underlying space is two and that the above matrix model in this phase is in the Ising universality class. The Wigner semi-circle law is determined by the free propagator and thus by the metric aspects of the underlying space \cite{Steinacker:2005wj,Nair:2011ux, Polychronakos:2013nca,Tekel:2013vz}.
The operator $ TrM TrM^3$ is the crucial ingredient in stabilizing the Ising phase. Indeed, given hindsight, a simpler set of parameters is given simply by  $C^{'}=-N$ while setting all the other parameters to zero. 

The configuration $M=a{\bf 1}_{2N}$, for our case (\ref{our_case}), is a solution of the equation of motion iff
\begin{eqnarray}
a&=&0 ({\rm disordered}),\nonumber\\
a^2&=&-\frac{\tilde{B}+1}{\sqrt{N}(2\tilde{C}-1)}({\rm uniform~ordered}).
\end{eqnarray}
The large $N$ scaling of the various coupling constants is given by 
\begin{eqnarray}
\tilde{B}=\frac{B}{N^{3/2}}~,~\tilde{C}=\frac{C}{N^2}.
\end{eqnarray}
Stability requires that $\tilde{B}\leq -1$ and $\tilde{C}\geq 1/2$. The configuration $M=a\gamma$ where $\gamma^2=1$ with $\gamma$ containing an equal numbers of $+1$ and $-1$ is also a solution iff
\begin{eqnarray}
a^2=-\frac{\tilde{B}}{\sqrt{N}(2\tilde{C}+3)}({\rm non~uniform~ordered}).
\end{eqnarray}
By substituting the configurations $M=a{\bf 1}_{2N}$ and $M=a\gamma$ in the Schwinger-Dyson identity (with $V=V_0+V_1$) 
\begin{eqnarray}
\frac{<V>}{N^2}=\frac{1}{4}+\frac{B}{2}\frac{<Tr M^2>}{N^2}+\frac{B^{'}}{2} \frac{<(TrM)^2>}{N^2},
\end{eqnarray}
we obtain the energies $E_1$ and $E_2$ in the two phases. 
It is not difficult to show that $E_1\leq E_2$ iff (recall that $\tilde{B}+1\leq 0$, $2\tilde{C}-1\geq 0$)
\begin{eqnarray}
\tilde{B}+1\leq -\frac{2\tilde{C}-1}{4}.
\end{eqnarray}
This result is confirmed non-perturbatively using Monte Carlo simulation of the eigenvalue problem corresponding to the diagonalization of the quartic multitrace matrix model $V=V_0+V_1$. Thus, there exists a region in the phase diagram in which the uniform ordered configuration is more stable than the disordered and the non-uniform ordered configurations \cite{Ydri:2015vba,Ydri:2015zsa}.
 
The partition function of the theory is given by
\begin{eqnarray}
Z=\int {\cal D} M \exp(-V[M]).
\end{eqnarray}
We will assume now that the matrix $M$ is $2N\times 2N$. Then, without any loss of generality, we can expand the matrix $M$ as
\begin{eqnarray}
M=M_0{\bf 1}_{2N}+M_1~,~Tr M_1=0.
\end{eqnarray}
Hence
\begin{eqnarray}
M_1=\sigma_aX_a~,~
M_0=a+m,
\end{eqnarray}
where $\sigma_a$ are the standard Pauli matrices, $m$ is the fluctuation in the zero mode, and $X_a$ are three hermitian $N\times N$ matrices. By substitution, we obtain immediately the model
\begin{eqnarray}
Z&=&\int {\cal D} X_a \exp(-V[X]))\int dm\exp(-f[m]).\label{fundamental3}
\end{eqnarray}
The potential $V$ is given now by the $SO(3)-$symmetric three matrix model 
\begin{eqnarray}
V&=&-CTr[X_a,X_b]^2+2CTr(X_a^2)^2+4D(Tr X_a^2)^2\nonumber\\
&+&2(B+\beta_0 a^2)TrX_a^2+2ia\gamma\epsilon_{abc} TrX_aX_bX_c.\label{fundamental2}
\end{eqnarray} 
The new coefficients $\beta_0$ and $\gamma$ are given below. 

We observe that the Chern-Simons term is proportional to the value $a$ of the order parameter. Thus, it is non-zero only in the Ising phase, and as a consequence, by tuning the parameters appropriately to the region in the phase diagram where the Ising phase exists, we will induce a non-zero value for the Chern-Simons. This is effectively the Myers term responsible for the condensation of the geometry \cite{Myers:1999ps,Azuma:2004zq}. The above multitrace three matrix model is then precisely a random matrix theory describing non-commutative gauge theory on the fuzzy sphere, where the first term is the Yang-Mills piece, whereas the second and fourth terms combine to give mass and linear terms for the normal scalar field on the sphere (recall that $a$ runs from $1$ to $3$). The third doubletrace term, proportional to $D$, depends only on the zero mode of the normal scalar field. Thus, it is not expected to play a major role in our discussion here. If we simply set $D=0$ then we get essentially the random matrix theory describing non-commutative gauge theory on the fuzzy sphere  found in   \cite{Steinacker:2003sd} (notice that the parameter $C$ appears in front of the Yang-Mills as well as in front of the normal mass term and thus can be scaled away). This theory itself is a generalization of the stringy non-commutative gauge theory on the fuzzy sphere considered in \cite{Alekseev:2000fd,Iso:2001mg,CarowWatamura:1998jn}.

Furthermore, we have shown recently in \cite{Ydri:2016osu} that the above model with $D=0$ sustains an absolutely stable emergent fuzzy two-sphere geometry, and as a consequence, the expansion around this emergent geometry to obtain a non-commutative gauge theory is fully consistent for all values of the gauge coupling constant. In other words, there is no phase transition to a Yang-Mills matrix phase at a finite value of the gauge coupling constant. This is expected to hold also for $D\neq 0$. 

However, we should emphasis here that we have obtained dynamically this slightly generalized version of non-commutative gauge theory on the fuzzy sphere by going to the phase of the model where a non-zero uniform order persists, and then by expanding around this order, we secure a non-zero Chern-Simons term crucial for the underlying emergent geometry of the fuzzy sphere. Recall that in \cite{Steinacker:2003sd}, this was achieved by constraining the matrix $M$ directly in a particular way. 
 
What is the effect of the zero mode $m$?

The potential $f$ in (\ref{fundamental3}) of the zero mode $m$ is given by
\begin{eqnarray}
f
&=&\big[2i\gamma\epsilon_{abc} TrX_aX_bX_c+4\alpha a^3+4\beta_0 aTrX_a^2+2a\beta_1\big]m\nonumber\\
&+&\big[2\beta_0Tr X_a^2+\beta_1+6\alpha a^2\big] m^2+4\alpha a m^3+\alpha m^4.
\end{eqnarray}
The various new coefficients, for our case (\ref{our_case}), are given by

\begin{eqnarray}
&&\beta_0=3N^2(2\tilde{C}-1)~,~\beta_1=2N^{5/2}(\tilde{B}+1)\nonumber\\
&&\alpha=N^3(2\tilde{C}-1)~,~\gamma=2N^2(2\tilde{C}-1).
\end{eqnarray}
The integration over $m$ can be done. The leading contribution in the large $N$ limit is essentially the one-loop result and it is by construction subleading compared to (\ref{fundamental2}). This integral consists of some function of $TrX_a^2$ and $i\epsilon_{abc}TrX_aX_bX_c$. The conclusion of the foregoing discussion remains practically unchanged with the addition of these multitrace corrections.   


Thus, the multitrace one matrix model (\ref{fundamental1}), which involves only a single hermitian matrix $M$ with $U(2N)$ symmetry, and which exhibits a uniform order for some values of the mass parameter $B$ and the quartic coupling constant $C$,  provides a new mechanism for emergent fuzzy two-sphere and its non-commutative gauge theory. Indeed, by expanding around the uniform order, and then performing the integral over the zero mode $m$, it is seen that the resulting theory given by the mostly (since $D\neq0$) single trace $SO(3)-$symmetric three matrix model (\ref{fundamental2}), plus  the multitrace corrections induced by the integral over $m$, is a matrix model describing a non-commutative gauge theory on the fuzzy sphere. The multitrace corrections are all subleading in $N$, with the exception of the term $(TrX_a^2)^2$ in the model  (\ref{fundamental2}),  and they depend generically on the mass deformations $TrX_a^2$ and $i\epsilon_{abc}TrX_aX_bX_c$. 

The model (\ref{fundamental2}) with $D=0$ has been shown in \cite{Ydri:2016osu} to sustain an absolutely stable emergent fuzzy sphere with fluctuation given by a non-commutative $U(1)$ gauge theory very weakly coupled to a scalar field (playing the role of dark energy). In the full multitrace model there exists always the possibility of a phase transition  from the fuzzy two-sphere phase to a Yang-Mills matrix phase \footnote{In Yang-Mills matrix models with a non-zero Chern-Simons term there exists typically a phase transition from the fuzzy sphere phase to a Yang-Mills matrix phase.} which occurs at some finite value of the gauge coupling constant. However, this is not expected since $(TrX_a^2)^2$ depends only on the zero mode of the normal scalar field while all other multitrace corrections are subleading in $N$. In other words, the full multitrace matrix model is also expected to sustain an absolutely stable emergent fuzzy sphere with fluctuation given by a slightly generalized non-commutative $U(1)$ gauge theory coupled to a scalar field.

In summary, the multitrace matrix model  (\ref{fundamental1}) gives a novel and a much simpler  alternative to Connes noncommutative geometry \cite{Connes:1996gi} and to the IKKT matrix model \cite{Ishibashi:1996xs} for emergent geometry in two dimensions.

\paragraph{Acknowledgments:}
This research was supported by CNEPRU: "The National (Algerian) Commission for the Evaluation of
University Research Projects"  under contract number ${\rm DO} 11 20 13 00 09$. We would like also to acknowledge generous funding and warm hospitality of DIAS (Dublin) and ICTP (Trieste).

\end{document}